\documentclass[aps,twocolumn,showpacs,preprintnumbers,floatfix]{revtex4}
\usepackage{graphicx}% Include figure files
\usepackage{epsfig}
\usepackage{dcolumn}% Align table columns on decimal point
\usepackage{epstopdf}
\usepackage{amsmath}
\newcommand{\Rmnum}[1]{\expandafter\@slowromancap\romannumeral #1@}
\begin{document}

\title{Anatomy of the $\rho$ resonance from lattice QCD at the physical point}
\author{\small 
Wei Sun,${}^{1,2}$ Andrei Alexandru,${}^3$ Ying Chen,${}^{1,2}$\footnote{cheny@ihep.ac.cn}  Terrence Draper,${}^4$ Zhaofeng Liu,${}^1$ and Yi-Bo Yang${}^{5}$\\
($\chi$QCD Collaboration)
}
\affiliation{ \small
$^1$ Institute of High Energy Physics, Chinese Academy of Sciences, Beijing 100049, China\\
$^2$ School of Physics, University of Chinese Academy of Sciences, Beijing 100049, China\\
$^3$ Department of Physics, George Washington University, Washington, DC 20052, USA\\
$^4$ Department of Physics and Astronomy, University of Kentucky, Lexington, KY 40506, USA\\
$^5$ Department of Physics and Astronomy, Michigan State University, East Lansing, MI 48824, USA}

\begin{abstract}
We propose a strategy to access the $q\bar{q}$ component of the $\rho$ resonance in lattice QCD. Through
a mixed action formalism (overlap valence on domain wall sea), the energy of the $q\bar{q}$ component is
derived at different valence quark masses, and shows a linear dependence on $m_\pi^2$. The slope is determined
to be $c_1=0.505(3)\,{\rm GeV}^{-1}$, from which the
valence $\pi \rho$ sigma term is extracted to be $\sigma_{\pi \rho}^{(\rm val)}=9.82(6)$ MeV using the
Feynman-Hellman theorem. At the physical pion mass, the mass of the $q\bar{q}$ component is interpolated to
be $m_\rho=775.9\pm 6.0\pm 1.8$ MeV, which is close to the $\rho$ resonance mass. We also obtain the leptonic decay
constant of the $q\bar{q}$ component to be $f_{\rho^-}=208.5\pm 5.5\pm 0.9$ MeV, which can be compared with the experimental value $f_{\rho}^{\rm exp}\approx 221$ MeV through the relation $f_{\rho}^{\rm exp}=\sqrt{Z_\rho}f_{\rho^\pm} $ with $Z_\rho\approx 1.13$ being the on-shell wavefunction renormalization of $\rho$ owing to the $\rho-\pi$ interaction. We emphasize that $m_\rho$ and $f_\rho$ of the $q\bar{q}$ component, which are
obtained for the first time from QCD, can be taken as the input parameters of $\rho$ in effective
field theory studies where $\rho$ acts as a fundamental degree of freedom.
\end{abstract}


\maketitle

\section{Introduction}

The vector meson $\rho$ is a well-known hadron resonance which appears in the $I=1$ and
$L=1$ $\pi\pi$ system with the resonance parameters $m_\rho=775$ MeV and $\Gamma_\rho=149$ MeV. On
the other hand, $\rho$ is assigned in the quark model to be the $I=1$ member of the $q\bar{q}$
vector meson nonet with mass around 1 GeV. The connection between the resonance $\rho$ in experiments
and the confined $q\bar{q}$ quark model $\rho$ was established by Jaffe by introducing concepts such
as {\it ordinary} and {\it extraordinary} hadrons ~\cite{Jaffe:2007}. In this picture, $q\bar{q}$ mesons,
glueballs, and hybrids are {\it ordinary} mesons, which decay into multi-hadron final states through the
creation of new quanta ($q\bar{q}$ pairs or gluons) and develop widths proportional to $1/N_c$. In the large
$N_c$ limit, ordinary mesons decouple and appear as bound states with discrete energies, such that
the Hilbert space is composed of discretized bound states and multi-hadron continuum states. In contrast,
{\it extraordinary} hadrons show up as resonances in hadron-hadron interactions but diminish in large $N_c$.
As far as the $\rho$ meson is concerned, there is a confined channel corresponding to the quark model $\rho$ and
its excited states, as well as an open channel of $\pi\pi$ scattering states. The coupling between both channels
results in the $\rho$ resonance of an $O(1/N_c)$ width. This argument is coincident
with the result of a chiral perturbation theory study of $\rho$ ~\cite{Pelaez:2003dy} that while the $\rho$
mass keeps almost constant, the width decreases with increasing $N_c$. This implies that $\rho$ is
a well-defined confined $q\bar{q}$ state in the $N_c\rightarrow \infty$ limit.

Even though this picture cannot be tested experimentally since $N_c=3$ in the real world and $\rho$ usually
shows up as a resonance, one can resort to the lattice QCD formalism for the related investigation. On the
finite Euclidean space-time lattice, the eigenstates of the QCD Hamiltonian have a discrete spectrum.
For the case of $\rho$, if there exists a Hilbert space expanded by both the non-interacting $\pi\pi$  states
and confined $q\bar{q}$ states, the QCD eigenstates can be viewed as state vectors in this Hilbert space. The last
decade witnessed extensive lattice QCD efforts on the $\rho$ resonance from $\pi\pi$ scattering~~\cite{Feng:2011rho,Pelissier:2012pi,Aoki:2011yj,Dudek:2012xn,Bali:2015gji,Guo:2016zos,
Fu:2016itp,Bulava:2016mks}, where the
use of $q\bar{q}$ operator and $\pi\pi$ operators is mandatory and the eigen energies are used to
extract the resonance parameters of $\rho$ using L\"{u}scher's formalism~\cite{Luscher:1986pf}.
In addition to the great success in this direction, it is also an interesting question whether the would-be confined
$q\bar{q}$ $\rho$ can
be accessed directly through the full-QCD lattice calculation. The major consideration is that one
can use an interpolation field operator which couples weakly to $\pi\pi$ states but which couples
almost exclusively to $q\bar{q}$ confined states. We find that the Coulomb gauge fixed wall-source
$q\bar{q}$ operator serves this goal. As such, we can obtain the mass of the $\rho$ bound state and
its decay constant, as well as the chiral behavior of these quantities. Phenomenologically, the
properties of the confined $q\bar{q}$ state $\rho$ may shed light on the intrinsic dynamics of the $\rho$
resonance. This strategy can be potentially extended to the study of other resonances, such as
$\Delta$ baryon, $K^*$ resonance etc.

This paper is organized as follows. Section 2 presents the derivation of the $m_\pi$ dependence of the confined $q\bar{q}$ $\rho$ mass and the relevant discussion. Section 3 is devoted to the extraction of the leptonic decay constant of $\rho$. The conclusions and a summary can be found in Section 4.

\section{$\rho$ meson mass}

Gauge configurations of $N_f=2+1$ domain-wall fermions with large spatial volume and physical pion
mass have been generated by the RBC \& UKQCD Collaborations~\cite{Blum:2014tka}. This work is based
on the 48I gauge ensemble with lattice size $L^3\times T=48^3\times 96$~\cite{Blum:2014tka}. The
lattice spacing has been determined to be $a^{-1}=1.730(4)$ GeV, such that the spatial extension of
the lattice is approximately $La\sim 5.5$ fm. The light sea quark mass is set to give the pion mass
$m_\pi^{(\rm sea)}=139.2(4)$ MeV. For the valence quarks, we adopt the overlap fermion action,
which is another realization of chiral fermions on the lattice. The low-energy constant
$\Delta_{\rm mix}$, which measures the mismatch of the mixed valence and sea pion masses between
the domain-wall fermion and the overlap fermion, is shown to be very small~\cite{Lujan:2012wg}.
Since overlap fermion accommodates the multi-mass algorithm and the eigenvectors are the same for
different quark masses, we use 1000 pairs of eigenvectors plus the zero modes for deflation in
calculating quark propagators for several masses on 45 configurations (see
Ref~\cite{Alexandru:2011sc} for details). The bare mass parameters are chosen as $am_q^{(\rm
val)}=0.00170, 0.00240, 0.00300, 0.00455, 0.00600$ and $0.02030$, which give the pion mass ranging
from $114$ to $371$ MeV. In this way we can discern the chiral behaviors of the mass and the
leptonic decay constant of the $\rho$ meson.

\begin{table*}[]
\caption{The table lists the pion masses $m_\pi$, the pion decay constants $f_\pi$, and the masses
of $\rho$ at different bare valence quark masses. \label{tab:mass_table} }
\begin{ruledtabular}
\begin{tabular}{ccccccc}
 $am_q^{(\rm val)}$ & 0.00170&0.00240&0.00300&0.00455&0.00600&0.02030\\
 $m_\pi$(MeV)  &  114(2)    & 135(2)   & 149(2)   & 182(2) & 208(2) &371(1)\\
 $f_\pi$(MeV)  &  130.3(9)  &131.0(9)  &131.6(8)) & ...    & ...       &...\\
 $m_\rho$(MeV) &  773(7)    & 775(6)   & 779(6)   & 784(5) & 789(5)  &836(3)\\
\end{tabular}
\end{ruledtabular}
\end{table*}

We first extract the decay constant of the pion according to the partially
conserved axial current relation
\begin{equation}\label{fpi}
m_\pi^2 f_\pi =(m_u+m_d)\langle 0|\bar{u}\gamma_5 d|\pi\rangle,
\end{equation}
which is free of renormalization since the quark mass renormalization constant $Z_m$ and the renormalization
constant $Z_P$ of the pseudoscalar density $\bar{u}\gamma_5 d$ satisfy the relation $Z_mZ_P=1$ for overlap
fermions. We obtain the pion masses and $\pi$ decay constants which are listed in
Table~\ref{tab:mass_table}. Through a linear interpolation in $m_\pi^2$
near the physical pion mass $m_\pi=139.5$ MeV, we get $f_\pi=131.3(6)$ MeV, which agrees with RBC\&UKQCD's result
$f_\pi=131.1(3)$ MeV on the same lattice and their final theoretical prediction $f_\pi=130.2(9)$ MeV~\cite{Blum:2014tka}.
RBC\&UKQCD also calculate $f_\pi$ on a larger lattice with a smaller lattice spacing, $a^{-1}=2.359(7)$ GeV,
with a result $f_\pi=130.9(4)$ MeV. Their $f_\pi$'s on the two lattices imply very small finite $a$
artifacts. This comparison can be taken as a calibration of our formalism.

In the calculation of the two-point functions in the $\rho$ channel, the quark propagators are
generated by spatial wall-sources after the gauge configurations are fixed to the Coulomb gauge
first. This corresponds to using the Coulomb gauge fixed wall-source operator for the charged
$\rho$,
\begin{equation}\label{wall-source}
O_{V,i}^{(\rm w)}(t)=\sum\limits_{\mathbf{y},\mathbf{z}}\bar{u}(\mathbf{y},t)\gamma_i
d(\mathbf{z},t).
\end{equation}
In principle, this operator couples to all the eigenstates of the lattice Hamiltonian, which can be
taken as the linear superpositions of $\pi\pi(I=1)$ scattering states and the confined $q\bar{q}$
states.
%In the center-of-mass (CM) frame of $\pi\pi$ system, the matrix element $\langle
%0|\bar{u}(\mathbf{x})\gamma_i d(\mathbf{-x})|\pi^-\pi^0;\mathbf{p}\rangle$ can be expressed as
%\begin{equation}
%\langle 0|\bar{u}(\mathbf{x})\gamma_i d(\mathbf{-x})|\pi^-\pi^0;\mathbf{p}\rangle \sim
%e^{i2\mathbf{p}\cdot \mathbf{x}} F(\mathbf{p}) p_i/L^3,
%\end{equation}
%where $\mathbf{p}$ is the spatial momentum of pions in the CM frame, say,
%$\mathbf{p}=\mathbf{p}_{\pi^-}=-\mathbf{p}_{\pi^0}$, and $F(\mathbf{p})$ is a scalar function
%depending on $\mathbf{p}$. As such the spatial summation in Eq.~(\ref{wall-source}) makes
%\begin{equation}
%\langle 0|O_{V,i}^{(\rm w)}|\pi^-\pi^0;\mathbf{p}\rangle\propto L^3 p_i
%F(\mathbf{0})\delta_{\mathbf{p},\mathbf{0}}.
%\end{equation}

\begin{figure}
\includegraphics[height=5cm]{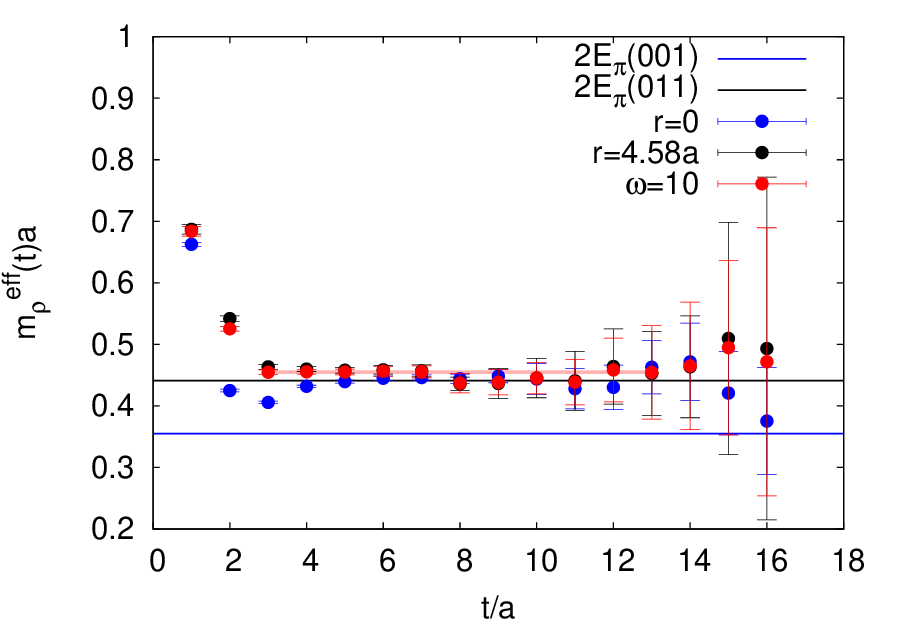}
\caption{The effective mass plateaus of $\rho$ at $m_\pi =208(2)$ MeV. The blue and black points
are from the correlation functions $C(r=0,t)$ and $C(r=4.58a,t)$, respectively. The red points are
from the mixed correlation function $C^{\omega}(t)=C(0,t)+\omega C(4.58a,t)$ with the mixing
parameter $\omega=10$. The red band shows the fitted mass in the time range $t/a \in [3,13]$. The
blue line and the black line are the energies of two non-interacting pions $2E_\pi$ with the
momentum modes $\mathbf{p}a=2\pi/L (0,0,\pm1)$ and $\mathbf{p}a=2\pi/L (0,\pm 1, \pm1)$,
respectively.\label{plat}}
\end{figure}

For the sink operators of the vector, we use the spatially extended operators $O_{V,i}(\mathbf{x},t;\mathbf{r})$
by splitting the quark and antiquark field operators with different spatial displacements
$\mathbf{r}$, namely, $O_{V,i}(\mathbf{x},t;\mathbf{r})=\bar{u}(\mathbf{x},t)\gamma_i
d(\mathbf{x}+\mathbf{r},t)$. Subsequently, the two-point functions $C(t,r)$ with different spatial
separation $r$ are calculated as
\begin{equation}\label{two-point}
C(r,t)=\frac{1}{3N_r}\sum\limits_{\mathbf{x},i, |\mathbf{r}|=r}\langle
0|O_{V,i}(\mathbf{x},t;\mathbf{r})O_{V,i}^{(\rm w) \dagger}(0)|0\rangle,
\end{equation}
where $N_r$ is the number of $\mathbf{r}$'s that satisfy $|\mathbf{r}|=r$.

The effective mass plateaus of $C(r,t)$ with $r=0$ (blue points) and
$r=\sqrt{20}a=4.58a$ (black points) at $m_\pi = 208(2)$ MeV are plotted in Fig.~\ref{plat}. It is
seen that the plateaus lie on top of each other in the large-time range. The
difference of the plateaus in the short-time range shows the $r$-dependence of the contamination
from higher states. In order to reduce the excited-state contamination, we linearly combine the two
correlation functions as $C_{\rm mix}(t)=C(0,t)+\omega C(4.58a,t)$ with an optimal mixing parameter
$\omega\approx 10$, by which we can get a very flat effective mass plateau starting from $t/a=3$, as shown
in the figure (red points). We fit $C_{\rm mix}(t)$ using a single-exponential form in the time
range $t/a\in[3,13]$ and get $m_Va = 0.456(4)$ (plotted as a red band), where the error is
statistical and is obtained through a jackknife analysis. We also plot the two lowest $\pi\pi$
$P$-wave thresholds $2E_\pi(001) = 614$ MeV (shown in Fig.~\ref{plat} as a blue line) and
$2E_\pi(011) = 761$ MeV (shown in Fig.~\ref{plat} as a black line) with the relative momenta
$\mathbf{p}a=2\pi/L (0,0,\pm1)$ and $\mathbf{p}a=2\pi/L (0,\pm1,\pm1)$, respectively. Since
$2E_\pi(001)$ is far from the expected $\rho$ mass, the corresponding $\pi\pi$ state should mix
little with $\rho$ and therefore have an energy close to $2E_\pi(001)$, but we do not observe this
state.

The disappearance of the $\pi\pi$ states can be tentatively understood as follows. Actually, the wall-source
operator $O_{V,i}^{(\rm w)}(t)$ can be re-expressed as
\begin{equation}
O_{V,i}^{(\rm w)}(t)=\sum\limits_{\mathbf{y},\mathbf{z}}\bar{u}(\mathbf{y},t)\gamma_i
d(\mathbf{z},t)\equiv\bar{\hat{u}}(\mathbf{0},t)\gamma_i\hat{d}(\mathbf{0},t),
\end{equation}
where $\hat{u}(\mathbf{0},t)$ and $\hat{d}(\mathbf{0},t)$ are the Fourier transformed quark fields
in the momentum space with the spatial momentum $\mathbf{q}=\mathbf{0}$. Qualitatively in the picture
of the non-relativistic constituent quark model, the matrix element
$\langle 0|\bar{\hat{u}}(\mathbf{0},t)\gamma_i\hat{d}(\mathbf{0},t)|\rho^{-}\rangle$ can be interpreted
as the probability amplitude of annihilating a zero-momentum anti-$u$ quark and a zero-momentum
$d$ quark in $\rho^{-}$ state. If $\rho^{-}$ is at rest, then the average momenta of the constituent quarks
are zero. Thus there is no suppression for this matrix element. However, for a $P$-wave $\pi^-\pi^0$
scattering state, the two pions must have non-zero relative momentum. In the center-of-mass frame of
the two pions, let the momenta of $\pi^-$ and $\pi^0$ be $\mathbf{p}$ and $-\mathbf{p}$, respectively.
Then the average momenta of the anti-$u$ quark in $\pi^-$ and the $d$ quark in $\pi^0$ are necessarily
non-zero. Therefore the matrix element
$\langle 0|\bar{\hat{u}}(\mathbf{0},t)\gamma_i\hat{d}(\mathbf{0},t)|\pi^{-}\pi^0\rangle$ will be strongly
suppressed. On the other hand, we assume the $q\bar{q}$ confined states and the non-interacting
$\pi\pi$ states establish a complete state basis for the Hilbert space when the $\rho-\pi\pi$ coupling is
switched off. After inserting these states, the correlation function
Eq.~\ref{two-point} can be expressed as
\begin{widetext}
\begin{eqnarray*}
C(r,t)&=&\frac{1}{3N_r}\sum\limits_{i, |\mathbf{r}|=r}\left[\langle
0|O_{V,i}(\mathbf{0},t;\mathbf{r})|\rho^-\rangle \frac{1}{2m_{\rho^-}V}\langle \rho^-|O_{V,i}^{(\rm w) \dagger}(0)|0\rangle\right. \nonumber\\
&+&\left. \sum\limits_{\mathbf{p}}\langle
0|O_{V,i}(\mathbf{0},t;\mathbf{r})|\pi^-(\mathbf{p})\pi^0(-\mathbf{p})\rangle \left(\frac{1}{2E_\pi(\mathbf{p})V}\right)^2
\langle \pi^-(\mathbf{p})\pi^0(-\mathbf{p})|O_{V,i}^{(\rm w) \dagger}(0)|0\rangle+\ldots \right]
\end{eqnarray*}
\end{widetext}
where $V=L^3 a^3$ is the spatial volume of the lattice, $1/(2m_{\rho^-}V)$ comes from the nomalization of $|\rho^-\rangle$,
and $(1/(2E_\pi(\mathbf{p})V))^2$ comes from the $\pi\pi$ state $|\pi^-(\mathbf{p})\pi^0(-\mathbf{p})\rangle$. So
the contribution of $\pi\pi$ states has an additional $1/L^3$ suppression factor.

This discussion also applies to the $\pi\pi$ state near the threshold $2E_\pi(011) = 761$ MeV. So we argue that the plateau comes predominantly
from the would-be $q\bar{q}$ confined state instead of the corresponding scattering state. In order to understand this theoretically,
let us consider a two-state system composed of the $q\bar{q}$ confined state $|\rho\rangle$ and the non-interacting
$\pi\pi$ state $|\pi\pi\rangle$ with
$H_0|\pi\pi\rangle=E_1|\pi\pi\rangle$, $H_0|\rho\rangle=E_2|\rho\rangle$, where $H_0$ is the
Hamiltonian without coupling between $\rho$ and $\pi\pi$. With the interaction of $\rho$ and
$\pi\pi$ included, the effective Hamiltonian can be written as the following $2\times 2$
matrix in the representation space spanned by $|\rho\rangle$ and $|\pi\pi\rangle$,
$H=H_0+H_{\rm I}=\left(
\begin{array}{cc}
  E_1 & x \\
  x & E_2
\end{array}
\right)$. Upon introducing the parameters $M=\frac{1}{2}(E_1+E_2)$ (we assume $E_2>E_1$),
$\Delta=E_2-E_1$ , and $\delta=\sqrt{1+4x^2/\Delta^2}$, the eigenvalues of $H$ are $E_\pm = M\pm
\frac{1}{2}\Delta \delta$, which satisfy $H|\alpha_\pm\rangle = E_\pm|\alpha_\pm\rangle$ with
$|\alpha_\pm\rangle=a_\pm |\pi\pi\rangle+b_\pm|\rho\rangle$.
The explicit expressions of $a_\pm$ and $b_\pm$ are
\begin{equation}
\left(\begin{array}{cc}
  a_- & b_- \\
  a_+ & b_+
\end{array}
\right)=\frac{1}{\sqrt{2\delta}} \left(\begin{array}{cc}
  \sqrt{\delta+1} & -\sqrt{\delta-1}  \\
  \sqrt{\delta-1}   & \sqrt{\delta+1}
\end{array}
\right).
\end{equation}
Subsequently, the Coulomb wall-source two-point function (note that the state normalization factor $1/2E$ has been absorbed into the
definition of $|\ldots\rangle$, since $\langle \ldots|\ldots\rangle =1$) can be expressed as
\begin{eqnarray}
C(t)&=&\langle O_P(t)O_W^{+}(0)\rangle\nonumber\\
&=&\langle 0|O_P|\alpha_-\rangle\langle\alpha_-|O_W^{+}|0\rangle e^{-E_-t}\nonumber\\
&+&\langle 0|O_P|\alpha_+\rangle\langle\alpha_+|O_W^{+}|0\rangle e^{-E_+ t}.
\end{eqnarray}
Applying the relation $\langle 0|O_W|\pi\pi\rangle=0$ and defining $Z_P=\langle 0|O_P|\rho\rangle$
and $Z_W=\langle 0|O_W|\rho\rangle$, the above equation can be rewritten as
\begin{eqnarray}
C(t)&=& Z_P Z_W (b_-^2 e^{-E_- t}+b_+^2 e^{-E_+ t})\nonumber\\
&=& Z_PZ_W e^{-(M+\frac{1}{2}\Delta)t}\left[1+\frac{1}{2}x^2 t^2\left(1+O(\Delta t)\right)\right]
\end{eqnarray}
where $M+\frac{1}{2}\Delta=E_2$ is exactly the mass of the $q\bar{q}$ confined state $\rho$
($H_0|\rho\rangle=E_2|\rho\rangle$) as defined before. One can also estimate $x$ as
follows~\cite{McNeile:2002fh}. According to Fermi's Golden Rule, the partial
decay width of $\rho\rightarrow \pi\pi$ is expressed as $\Gamma=2\pi\langle x^2\rangle \rho(E)$
(note that $x=\langle \rho|H_I|\pi\pi\rangle$),
where the angle bracket means the average over the spatial angle with $\langle x^2\rangle=x^2/3$,
and $\rho(E)=L^3 kE/(16\pi^2)$ is the spectral density. Thus we have $\Gamma=x^2L^3kE/(24\pi)$,
which gives an estimate $ax\sim 0.025$ using the physical width $\Gamma_\rho\sim 150$ MeV and
$a^{-1}=1.73$ GeV. If one uses the single-exponential function to fit the correlation function, the
contribution of the $x^2$ term will give roughly $\le 1\%$ relative deviation from $E_2$, which
is much smaller than the statistical errors and negligible. Thus we have argued that the plateau
corresponds to the mass of the $q\bar{q}$ confined state $\rho$
\begin{equation}
C(t)\approx Z_P Z_W e^{-E_2 t}.
\end{equation}

\begin{figure}
\includegraphics[width=8.0cm]{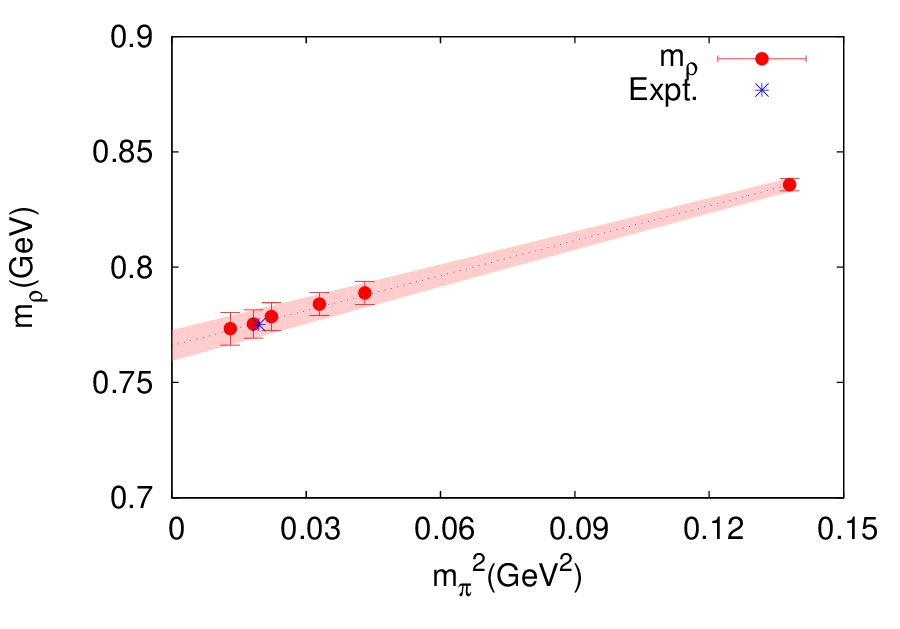}
\caption{ The $\rho$ mass $m_\rho=775.9\pm 6.0$ MeV at the physical point is interpolated by
$m_\rho(m_\pi)=m_\rho(0)+c_1 m_\pi^2$~\cite{Bruns:2005}. The red band shows the error of the interpolation. The
experimental value of $m_\rho=775$ MeV is also plotted as the blue cross for comparison.\label{mass_chiral}}
\end{figure}

We take a similar analysis procedure for the correlation functions at other pion masses, and the
extracted masses of $\rho$ are listed in Table~\ref{tab:mass_table} and are also plotted in
Fig.~\ref{mass_chiral} with respect to $m_\pi^2$, by which the chiral behavior of $m_\rho$ can be
investigated. From Fig.~\ref{mass_chiral} it is seen that $m_\rho$ is very linear in $m_\pi^2$ for
$m_\pi$ ranging from 114 MeV to 371 MeV. A correlated jackknife analysis using the form~\cite{Bruns:2005}
\begin{equation} \label{vmass_fit}
m_\rho(m_\pi)=m_\rho(0)+c_1 m_\pi^2,
\end{equation}
gives $m_\rho(0)=766(7)$ MeV and $c_1=0.505(3)\,{\rm GeV}^{-1}$ with $\chi^2/d.o.f$ = 0.13. The
$\rho$ mass at the physical $m_\pi$ is $m_\rho=775.9\pm 6.0\pm 1.8$ MeV, where the second error is
due to the 0.23\% uncertainty of the lattice spacing. Our data cannot discern higher order terms in
$m_\pi$. We would like to point out that our study is carried out for the first time in the chiral
region around the physical point and with chiral fermions, although there have been many lattice
studies on this
topic~\cite{Bernard:2001av,Leinweber:2001ac,Allton:2005fb,Armour:2005mk}. We note that
$c_1$ is precisely determined, and serves potentially as a constraint on the chiral perturbation
study of $\rho$. Furthermore, $c_1$ is exactly the valence or connected insertion part of the $\pi
\rho$ sigma term from the Feynman-Hellman theorem \mbox{$\sigma_{\pi \rho}^{(\rm val)} =
m_{\pi}^2\, dm_{\rho}/dm_{\pi}^2 = c_1 m_\pi^2$}, since the sea is fixed in our partially quenched
calculation of $m_{\rho}$. From the fitted $c_1$ in Eq.~(\ref{vmass_fit}), we find $\sigma_{\pi
\rho}^{(\rm val)} = 9.82(6)$ MeV. One can determine the disconnected part from a direct calculation
of the $m \bar{\psi}\psi$ matrix element in the disconnected three-point correlator.
\par

\section{The leptonic decay constant of the $\rho$ meson}
The calculation of the decay constant of the charged $\rho$ is straightforward~\cite{Lewis:1996qv,AliKhan:2001tx,Gockeler:2005mh,Hashimoto:2008xg,McNeile:2002fh,Jansen:2009hr}. For the charged $\rho$, for
example, $\rho^-$, $f_{\rho^-}$ is defined by
\begin{equation}  \label{rho_decay_constant}
 \langle 0|J_\mu^{(-)}(0)|\rho^-(\vec{p},\zeta)\rangle = m_{\rho} f_{\rho^-} \epsilon_\mu(\vec{p},\zeta),
\end{equation}
where $J_\mu^{(-)}(x)=(\bar{u}\gamma_\mu d)(x)$ is the charged vector current and
$\epsilon_\mu(\vec{p},\zeta)$ is the $\zeta$-th polarization vector of $\rho^-$ with $\zeta=1,2,3$.
The spatial components of $J_\mu^{(-)}(x)$ are actually the operators $O_{V,i}(x;\mathbf{r}=0)$;
therefore, the matrix element defined in Eq.~(\ref{rho_decay_constant}) can be extracted from
$C(r=0,t)$. The key challenge is to divide out the matrix element of the wall source operator
$\langle 0|O_{V,i}^{(\rm w)}|V(\vec{p},\zeta)\rangle$. Usually this matrix element can be derived
by calculating the wall-wall correlation function
\begin{eqnarray}\label{wall}
C^{(\rm w)}(t)&\equiv& \sum\limits_{r} N_r C(r,t)=\frac{1}{3} \sum\limits_i\langle
0|O_{V,i}^{(\rm w)}(t)O_{V,i}^{(\rm w),\dagger}(0)|0\rangle\nonumber\\
&=&\frac{1}{3}\sum\limits_{\mathbf{x,r},i}\langle
0|O_{V,i}(\mathbf{x},t;\mathbf{r})O_{V,i}^{(\rm w),\dagger}(0)|0\rangle,
\end{eqnarray}
where the last equation uses the definition of $C(r,t)$ in Eq.~(\ref{two-point}).

%However, $C^{(\mathrm w)}(t)$ is usually very noisy owing to the rapid damping of $C(r,t)$ with %increasing
%$r$. In order to increase the signal-to-noise ratio of $C^{(\mathrm w)}(t)$, we adopt an
%approximation
%\begin{equation}
%C^{(\rm w)}(r_c,t)=\sum\limits_{r\le r_c}N_rC(r,t),
%\end{equation}
%with the truncation size $r_c$ large enough to saturate $C^{(\mathrm w)}(t)$. We observe that the
%ground state contribution to $C^{(\mathrm w)}(t)$ decays as $\sim e^{-({\frac{r}{r_0}})^\kappa}$
%with the parameters $\kappa\sim 1.60$ and $r_0\sim 6a$, which are insensitive to $m_\pi$.

\begin{figure}
\includegraphics[width=8.0cm]{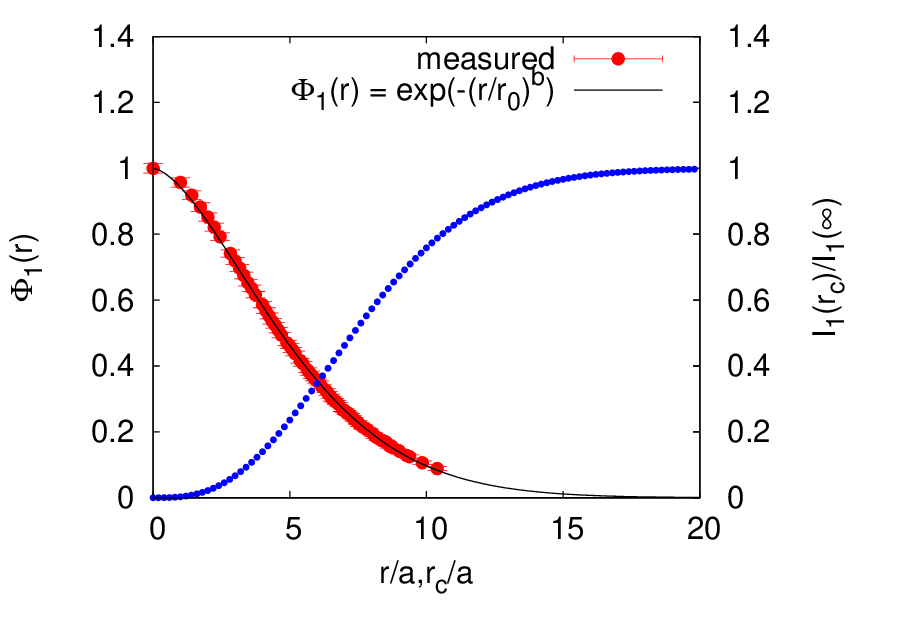}
\caption{The red points show $\Phi_i(r)$ which depicts the fall-off of $C(r,t)$ when $r$
increases. The parameterization of $\Phi_1(r)$ is plotted by the curve. The blue points are the
ratios $I_1(r_c)/I(\infty)$ at different $r_c$. \label{wavefun}}
\end{figure}

However, a very
large statistics is required to obtain a satisfactory signal-to-noise ratio for this kind of
correlation function. The reason for noisy $C^{(\rm w)}(t)$ is explained as follows. Using the
spectral expression $C(r,t)=\sum\limits_i \Phi_n(r) e^{-E_n t}$, when $t\rightarrow\infty$ one has
\begin{eqnarray}
C^{(\rm w)}(t)&\approx& \sum\limits_r N_r \Phi_1(r) e^{-E_1 t}.
\end{eqnarray}
In practice, we calculate $C(r,t)$ for $r$ ranging from 0 to $10a$ and observe the profile of
$\Phi_1(r)$ for $t=7a$ where all $C(r,t)$ are almost saturated by the ground state. The $\Phi_1(r)$
at $m_\pi=208(2)$ MeV (normalized as $\Phi_1(0)=1$) is plotted in Fig.~\ref{wavefun} as red points.
It is seen that $\Phi_1(r)$ damps rapidly with $r$ and can be parameterized as
\begin{equation}\label{parameterization}
\Phi_1(r)=\Phi_1(0)e^{-\left({\frac{r}{r_0}}\right)^b},
\end{equation}
with the parameters $b=1.60$ and $r_0=5.88 a$. The curve illustrates this parameterization in
the figure.

We check this at other pion masses and find $\Phi_1(r)$ is similar for all the cases
and is very insensitive to $m_\pi$. This means that when calculating the wall-to-wall correlation
function $C^{(\rm w)}(t)$, the $C(r,t)$'s (see in Eq.~(\ref{wall})) with very large $r$ contribute
only noise and make $C^{(\rm w)}(t)$ very noisy. In order to circumvent this difficulty, we
introduce a cutoff $r_c$ to exclude the contributions of $C(r,t)$'s with $r>r_c$ from $C^{(\rm w)}(t)$
and use the correlation function~\cite{Liu:2017man}
\begin{equation}
C^{(\rm w)}(r_c,t)=\sum\limits_{r\le r_c}N_rC(r,t),
\end{equation}
to approximate $C^{(\rm w)}(t)$. Letting $I_1(r')=\int_0^{r'} dr r^2 \Phi_1(r)$, one can see that
the ratio $C^{(\rm w)}(r_c,t)/C^{(\rm w)}(t)$ can be depicted by the ratio $I_1(r_c)/I_1(\infty)$
at large $t$. The ratio $I_1(r_c)/I_1(\infty)$ using the parameterization above is also plotted in
Fig.~\ref{wavefun}.
%### Question: why there is no errors on the saturation curve?
It approaches to 1 beyond $r_c=15a$ and is equal to 0.995 at $r_c=20a$, whose deviation from one is
already much smaller than the statistical error. So we take $C^{(\rm w)}(20a,t)$ as a satisfactory
approximation of $C^{(\rm w)}$ throughout this work.
Using these parameters, the
theoretical ratio $C^{(\rm w)}(r_c,t)/C^{(\rm w)}(t)$ deviates from unity by roughly $0.5\%$ at
$r_c=20a$ in the time range where the ground state dominates, which is much smaller that the
relative error of $C^{(\mathrm w)}(r_c,t)$. So we take $r_c=20a$ in practice. With this
prescription, we jointly fit the following functions to extract the decay constant,
\begin{eqnarray}\label{jointfit}
C(0,t)&=&\sum\limits_{n} 2m_n L^3 f_{n}Z_n^{(\rm w)}e^{-m_n t}\nonumber\\
C^{(\rm w)}(20a,t)&\approx&\sum\limits_n 2m_n L^3 (Z_n^{(\rm w)})^2 e^{-m_n t},
\end{eqnarray}
where $Z_n^{(\rm w)}$ is the matrix element of the wall source operator between the vacuum and the
$n$-th state, and $f_n$ is the decay constant of the $n$-th state according to the definition
in Eq.~(\ref{rho_decay_constant}). In practice, two exponentials are used in the fit and $f_1$ is
taken as the bare decay constant of $\rho$. (The second term is introduced to account for the
contamination of higher states.) Since $f_1$ is sensitive to the value of $m_1$, we adopt the
single-elimination jackknife analysis procedure as follows. On each jackknife re-sampled ensemble,
we first obtain the mass parameter $m_1$ from $C_{\rm mix}(t)$ defined previously, and then extract
$f_1$ through a joint fit to Eq.~(\ref{jointfit}) with $m_1$ fixed. After that, we quote the
jackknife error of $f_1$ as the statistical error.

\begin{figure}
\includegraphics[width=8.0cm]{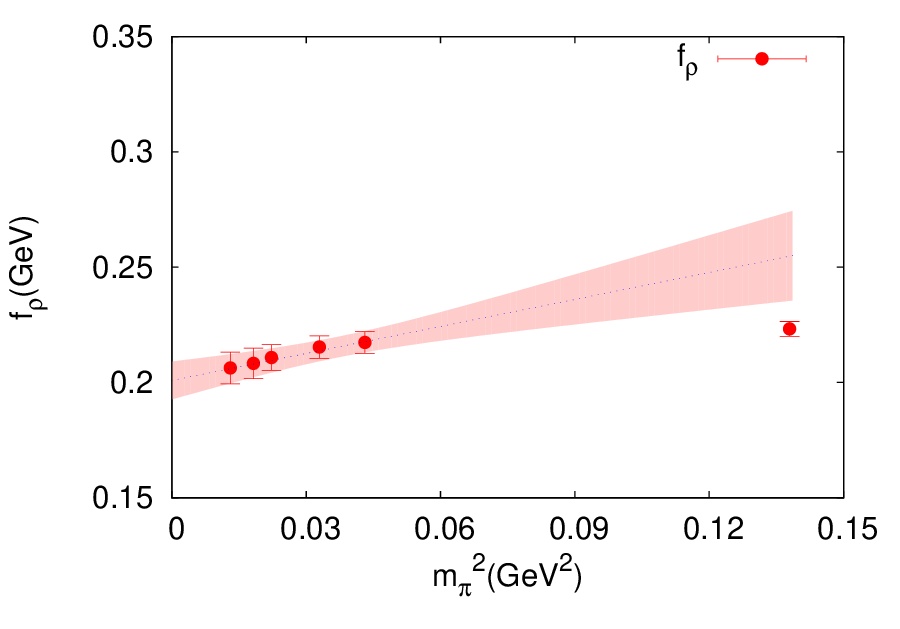}
\caption{ The decay constant $f_\rho$ obtained at different pion masses. The curve shows the linear interpolation in terms of $m_\pi^2$.\label{frho}}
\end{figure}

\begin{table*}[]
\caption{The renormalization constant $Z_A$ obtained at different pion masses, which also gives
$Z_V$ by the relation $Z_A=Z_V$ for overlap fermions. The renormalized decay constants $f_\rho$ are
also listed in the table. \label{fv} }
\begin{ruledtabular}
\begin{tabular}{lcccccc}
%% $am_q$  & 0.00170  & 0.00240 & 0.00300 & 0.00455 & 0.00600  & 0.02030\\
 $m_\pi$({\small MeV})  &  114(2)  & 135(2) & 149(2) & 182(2) & 208(2)& 371(1)\\
 $Z_A$    &  1.103(4)  & 1.103(3) & 1.104(2) & 1.104(2) & 1.105(1)&1.105(1) \\
 $f_\rho$({\small MeV})  &  206(7)  & 208(7) & 211(6) & 215(5) & 217(5)& 223(3)
\end{tabular}
\end{ruledtabular}
\end{table*}

In calculating the renormalization constant $Z_V$ of the vector current, we use
the relation $Z_V = Z_A$ ($Z_A$ is the renormalization constant of the axial vector current)
since the overlap fermions obey exact chiral symmetry on lattice. Following
the non-perturbative renormalization procedure in Ref.~\cite{Liu:2013yxz},
we calculate $Z_A$ from the Ward identity for a few bare quark masses which gives $Z_A=1.1045(8)$ in the
chiral limit. $Z_A$ and the renormalized
decay constant $f_\rho$ at different $m_\pi$ are listed in Table~\ref{fv}. Figure~\ref{frho} shows the
chiral behavior of $f_\rho$, from which we get
\begin{equation}\label{prediction}
f_{\rho^{\pm}}=208.5\pm 5.5\pm 0.9~\,{\rm MeV}
\end{equation}
through a linear interpolation in $m_\pi^2$ in the neighborhood of the physical pion mass, or
specifically, in the range $m_\pi^2 \in [0.012,0.044]\,{\rm GeV^2}$. We do not include the
result at $m_\pi=0.371$ GeV for the interpolation since the linear fit in $m_\pi^2$ is invalid
at this $m_\pi$.
The first error is statistical and the second is the combined uncertainty of $Z_V$, the scale
parameter $a^{-1}$, and the approximated wall-wall correlation function.

In $\tau$ decays, the branching fraction of the process $\tau\rightarrow \rho^- \nu_\tau$ is
$\mathcal{B}_{\rho}=25.21(33)\%$, which results from subtracting the 0.31(32)\% non-$\rho(770)$ contribution
from the $\tau\rightarrow \pi^-\pi^0 \nu_\tau$ branching fraction 25.52(9)\%)~\cite{PDG:2016}.
$\mathcal{B}_{\rho}$ gives $f_{\rho^-}^{\rm exp}=221.1\pm 1.6$ MeV when $\Gamma_\rho$ is taken into account.
On the other hand, the partial decay width $\Gamma(\rho^0\rightarrow e^+e^-)=7.04(6)$ keV~\cite{PDG:2016} gives
$f_{\rho^0}^{\rm exp}=221(1)$ MeV if one takes the quark-model value $\bar{Q}_{\rho^0}^2=1/2$ of $\rho^0$ effective
charge squared. $f_{\rho^0}^{\rm exp}\approx f_{\rho^\pm}^{\rm exp}$ is the natural result of the conservation of
the vector current (CVC). Obviously the experimental values deviate from our prediction by roughly 6\%. This discrepancy can be
understood as follows. If the $q\bar{q}$ confined state $\rho$ we have obtained is viewed as the free (bare) $\rho$ state,
$f_{\rho}$ is actually the transition amplitude of the free $\rho$ to a gauge boson ($W^\pm$ for charged $\rho$ and
photon for the neutral $\rho$). When the $\rho-\pi$ interaction is switched on, according to the LSZ reduction
formula, the amplitude $f_{\rho}^{\rm exp}$ of the physical $\rho$
is related to $f_{\rho}$ by $f_{\rho}^{\rm exp}=\sqrt{Z_\rho} f_\rho$ with $Z_\rho\approx 1.13$~\cite{Jegerlenhner:2011ti} the
on-shell wave function renormalization coming from the $\rho$ self-energy. In other words, the 6\% deviation from $f_\rho^{\rm exp}$
is exactly described by $\sqrt{Z_\rho}\approx 1.06$.

%$$\Gamma_{e^+e^-}=\frac{4\pi}{3}\alpha_{\mathrm QED}^2\bar{Q}_{\rho^0}^2\frac{f_{\rho^0}^2}{m_{\rho^0}}$$
%with $\alpha_{\rm QED}=1/137$ and
%$\bar{Q}_{\rho^0}^2=1/2$ the $\rho^0$ effective electric charge squared expected from the constituent quark model.

\section{Conclusion}

To summarize, we argue that the $q\bar{q}$ wall source operator in a fixed gauge can strongly suppress
$P$-wave scattering states such that the $q\bar{q}$ component of an ordinary meson, such as $\rho$, can
be accessed in lattice QCD study. For the case of $\rho$, this $q\bar{q}$ component can be taken as the bare $q\bar{q}$ confined
state with a mass of $m_\rho=775.9\pm 6.0\pm 1.8$ MeV at the physical
pion mass, which is almost the same as the pole mass of the $\rho$ resonance. This observation reinforces the chiral perturbation
theory study and the so-called {\it ordinary meson} argument of the $\rho$ resonance that the $\rho-\pi$ interaction does not
shift the mass of $\rho$ much but contributes to its width. $m_\rho$ is almost linear in $m_{\pi}^2$ and the slope
$c_1=0.505(3)\,{\rm GeV}^{-1}$ also gives the valence $\pi \rho$ sigma term which gives $\sigma_{\pi \rho}^{(\rm val)}= 9.82(6)$ MeV from
the Feynman-Hellman theorem. We also extract the leptonic decay
constant of the bare $\rho^\pm$ state to be $f_{\rho^{\pm}}=208.5\pm 5.5\pm0.9$ MeV at the physical $m_\pi$, whose deviation from
the experimental value $f_{\rho}^{\rm exp}\approx 221$ MeV is explained by $f_{\rho}^{\rm exp}=f_{\rho^\pm} \sqrt{Z_\rho}$ with $Z_\rho\approx 1.13$
being the on-shell wavefunction renormalization of $\rho$ owing to the $\rho-\pi$ interaction. This study may shed new light on the
nature of the $\rho$ resonance and also be helpful to understand the properties of other hadron resonances.

\section*{ACKNOWLEDGEMENT}
We are very grateful to the valuable discussion with Prof. Q. Zhao and Prof. K.-F. Liu. We thank the RBC\&UKQCD Collaborations for providing us their DWF gauge
configurations. Ths work is
supported in part by the U.S. DOE Grant No.\ DE-SC0013065, and also by the National Nature
Science Foundation of China (NSFC) under Grants No.~11335001, No.~11575196, No.~11575197 and No.~11621131001 (CRC110
by DFG and NSFC). A.A. is supported in part by the National Science Foundation CAREER grant PHY-1151648 and by U.S. DOE Grant No. DE-FG02-95ER40907.
Y.~C.\ thanks the support by the CAS Center for Excellence in Particle Physics
(CCEPP). This research used resources of the Oak Ridge Leadership Computing Facility at the Oak
Ridge National Laboratory, which is supported by the Office of Science of the U.S. Department of
Energy under Contract No.~DE-AC05-00OR22725.

\end{document}